\pdfoutput=1
\documentclass[superscriptaddress,twocolumn,aps,preprintnumbers,amsmath,amssymb,nofootinbib]{revtex4}

\usepackage{amsmath}
\usepackage{amssymb}
\usepackage{amsfonts}
\usepackage{graphicx}
\usepackage[dvipsnames]{xcolor}
\usepackage{xfrac}
\usepackage{comment}
\usepackage{pifont}
\usepackage{physics}
\usepackage{fourier}
\usepackage{hyperref}
\usepackage{bm}
\usepackage{enumitem}
\usepackage{makecell}

\usepackage{float}

\usepackage{tikz} % for drawing diagrams
\usetikzlibrary{positioning}
\usetikzlibrary{matrix}
\usetikzlibrary{arrows.meta}

\definecolor{rossoferrari}{HTML}{D9073D}
\definecolor{mediumblue}{HTML}{0000CD}
\definecolor{forestgreen}{HTML}{228B22}
\definecolor{desy_blue}{HTML}{009EE2}
\definecolor{desy_orange}{HTML}{FD8800}
\definecolor{light_pink}{rgb}{1,0.4,0.4}
\definecolor{light_blue}{rgb}{0.284602,0.317763,0.963947}
\hypersetup{setpagesize=false,bookmarksnumbered=true,bookmarksopen=true,%
colorlinks=true,linkcolor=light_blue,urlcolor=rossoferrari,citecolor=rossoferrari,linktocpage=false}

\newcommand{\xmark}{\ding{55}}

\begin{document}

%%%%%%%%%%%%%%%%%%%%%%%%%%%%%%%%%%%%%%%%%%%%%%%%%%%%%%%%%%%%%%%%%%%%%%%%%%%%%%%%%%%%%%%%%%%%%%%%%%%%

\preprint{MITP-25-051}

\title{Asymgenesis}

\author{Martin A.\ Mojahed}
\email{mojahedm@uni-mainz.de}
\affiliation{Physics Department T70, Technical University of Munich, 85748 Garching, Germany}
\affiliation{PRISMA$^+$ Cluster of Excellence \& Mainz Institute for Theoretical Physics, FB 08 - Physics, Mathematics and Computer Science, Johannes Gutenberg-Universität Mainz, Staudingerweg 9, 55099 Mainz, Germany}

\author{Sascha Weber}
\email{wesascha@uni-mainz.de}
\affiliation{PRISMA$^+$ Cluster of Excellence \& Mainz Institute for Theoretical Physics, FB 08 - Physics, Mathematics and Computer Science, Johannes Gutenberg-Universität Mainz, Staudingerweg 9, 55099 Mainz, Germany}

%%%%%%%%%%%%%%%%%%%%%%%%%%%%%%%%%%%%%%%%%%%%%%%%%%%%%%%%%%%%%%%%%%%%%%%%%%%%%%%%%%%%%%%%%%%%%%%%%%%%

\begin{abstract}
We present a framework based on the standard type-I seesaw model that relates the baryon asymmetry of the universe to the dark matter (DM) density. The framework, which we name \textit{Asymgenesis}, relies on the presence of primordial charge asymmetries seeded either in the dark sector or in the visible sector. A higher-dimensional portal operator reshuffles this initial asymmetry into both sectors, eventually resulting in a nonzero $B-L$ asymmetry and an asymmetric DM component. Compared to conventional asymmetric-dark-matter (ADM) schemes, our framework imposes far milder requirements on the portal interaction. In particular, the portal interaction need not violate $B-L$, and the temperature scales of efficient $B-L$ violation and efficient charge-transfer interaction mediated by the portal operator can be separated. We develop the formalism in detail and argue that the flexibility of our framework enlarges the model-building landscape for ADM. 

\end{abstract}

%%%%%%%%%%%%%%%%%%%%%%%%%%%%%%%%%%%%%%%%%%%%%%%%%%%%%%%%%%%%%%%%%%%%%%%%%%%%%%%%%%%%%%%%%%%%%%%%%%%%

\date{\today}
\maketitle

%%%%%%%%%%%%%%%%%%%%%%%%%%%%%%%%%%%%%%%%%%%%%%%%%%%%%%%%%%%%%%%%%%%%%%%%%%%%%%%%%%%%%%%%%%%%%%%%%%%%

\noindent\textbf{Introduction}\,---\, 
Addressing the observed baryon asymmetry of the Universe (BAU)~\cite{Planck:2018vyg,ParticleDataGroup:2024cfk} calls for physics beyond the Standard Model (SM). A leading paradigm for dynamically generating the BAU in the early universe is leptogenesis (LG)~\cite{Fukugita:1986hr}, which requires extending the SM field content by at least two right-handed neutrinos (RHNs). An attractive feature of LG is that RHNs can account for the observed nonzero masses of active neutrinos via the type-I seesaw mechanism~\cite{Minkowski:1977sc,Yanagida:1979as,Yanagida:1980xy,Gell-Mann:1979vob,Mohapatra:1979ia}, thereby connecting the origin of the BAU to the origin of active neutrino masses~\cite{Buchmuller:2005eh,Bodeker:2020ghk}. Most LG realizations rely on all three of Sakharov's conditions~\cite{Sakharov:1967dj} for generating the BAU to emerge
from the RHN sector~\cite{Fukugita:1986hr,Luty:1992un,Covi:1996wh,Pilaftsis:1997jf,Akhmedov:1998qx,Barbieri:1999ma,Pilaftsis:2003gt,Giudice:2003jh,Buchmuller:2004nz,Pilaftsis:2005rv,Klaric:2020phc,Klaric:2021cpi}. A notable exception is wash-in leptogenesis (WILG), which does not require any source of charge-parity $(CP)$-violation from the RHN sector~\cite{Domcke:2020quw, Domcke:2022kfs, Mojahed:2025vgf,WILGRelated}.

The philosophy underlying WILG is to treat RHN interactions as spectator processes whose role is to provide a source of $B\!-\!L$ violation. In this way, RHNs are treated on par with electroweak sphalerons~\cite{Kuzmin:1985mm}, whose role in standard LG mechanisms is merely to serve as a source of $B+L$ violation. Meanwhile, $CP$ violation is attributed to novel $CP$-violating dynamics at high temperatures above the mass of the heaviest RHN $T\gg M_N$, in a process referred to as chargegenesis (CG), see Refs.~\cite{Domcke:2022kfs,Schmitz:2023pfy,Mukaida:2024eqi,Mojahed:2024yus} for concrete realizations. The role of CG is to provide $CP$-violating initial conditions for WILG, and can be viewed as a UV completion for the effective framework of WILG. In particular, the SM entails eleven linearly independent conserved global charges at high temperatures that could be produced during CG~\cite{Domcke:2020quw}. During WILG, RHNs act on the nontrivial chemical background induced by CG in a $B-L$ violating manner and drive the thermal plasma to an attractor solution featuring nonzero $B-L$, even if $B-L$ was conserved throughout CG~\cite{Domcke:2020quw}. 

Within the framework presented in this paper, we demonstrate that the initial conditions required for successful WILG can also be used to address the generation of the DM in our universe~\cite{Planck:2018vyg,Cirelli:2024ssz}. In particular, if DM has a particle nature and consists primarily of either dark particles or dark anti-particles, then a primordial charge asymmetry in the SM sector can induce suitable initial conditions for ADM. Similarly, a primordial charge asymmetry in the dark sector can also induce initial conditions suitable for WILG. This idea is closely related to the standard ADM paradigm~\cite{Kaplan:1991ah,Nussinov:1985xr,Kaplan:2009ag,Petraki:2013wwa,Zurek:2013wia}, where the generation of the BAU and the DM abundance are tightly connected. In many ADM models, either a $B-L$ asymmetry or a DM asymmetry is generated at high temperatures, and subsequently transferred to the other sector~\cite{Shelton:2010ta,Haba:2010bm,Buckley:2010ui,Blennow:2010qp,Bernal:2016gfn,Hall:2019ank,Hall:2021zsk,Barman:2021tgt,Asadi:2025vli}.\footnote{See e.g. also~\cite{Mahapatra:2023dbr,Borah:2024wos,Takahashi:2026ngu} for some recent work on ADM in seesaw models.} A standard transfer mechanism is provided by a higher-dimensional operator of the form $\mathcal{O}_{B-L}\mathcal{O}_D/\Lambda^n$, where $\mathcal{O}_{B-L}$ violates $B-L$ and $\mathcal{O}_D$ violates a global DM charge~\cite{Zurek:2013wia}. At high temperatures, the interaction rates induced by this operator are efficient, and the asymmetry is freely shared between the two sectors. As the temperature in the early universe drops, the interaction decouples, and an asymmetry is separately frozen in the dark and visible sectors. This standard picture of charge transfer via higher-dimensional operators is illustrated on the left in Fig.~\ref{fig:ADM_Asymgenesis}. Finally, as the temperature drops below the mass of the DM particle, particle-antiparticle annihilations erase the symmetric component of the DM density, resulting in a DM density determined by the particle-antiparticle asymmetry and prediction of a $\mathcal{O}$(GeV) DM mass~\cite{Zurek:2013wia}. 

\begin{figure*}[t]
\centering
\begin{tikzpicture}[
  scale=0.8,
  transform shape,
  >=stealth,
  every node/.style={font=\normalsize}
]%[scale=0.8, transform shape, >=stealth]

%==================== LEFT: ADM ====================
\begin{scope}[]

% ---- ORIGINAL ADM CODE (T-axis REMOVED) ----
\begin{scope}[scale=1.2]

\draw[domain=-1.185:1.185, smooth, thick]
  plot (\x, {-2*(\x)^2 + 2});

\draw[->,thick] (-3.83,-0.5) -- (-3.83,3.5) node[left] {$T$};
\draw[thick] (-3.93,2) -- (-3.73,2);
\node[right] at (-3.73,2) {$T_{\mathrm{eq}}$};

\node (Lagrangian) at (0,2.5)
 {$\displaystyle \mathcal{L} \supset \frac{\mathcal{O}_{B\!-\!L}\mathcal{O}_{X}}{\Lambda^n}$};
\node (Transfer) [left=0.25cm of Lagrangian] {Charge transfer};
\node [draw,above=0cm of Lagrangian] {$B\!-\!L$ violation};

\node (SM Operator) at (-2.1,1)
 {$\displaystyle \mathcal{O}_{B-L} = \Bar{L}H,\dots$};
\node (SM text) [below=0cm of SM Operator] {SM sector};
\node [below=0cm of SM text] {$\mu_{B-L} \neq 0$};

\node (DM Operator) at (2,1)
 {$\displaystyle \mathcal{O}_X = X^p,\dots$};
\node (DM text) [below=0cm of DM Operator] {DM sector};
\node [below=0cm of DM text] {$\mu_X \neq 0$};

\end{scope}

% ---- Square bracket + label (ADM) ----
\draw[line width=0.8pt] (-4.6,-1.0) -- (-4.6,-0.5);
\draw[line width=0.8pt] ( 4.2,-1.0) -- ( 4.2,-0.5);
\draw[line width=0.8pt] (-4.6,-1.0) -- ( 4.2,-1.0);
\node at (0,-2.0) {\large\bfseries Standard ADM};

\end{scope}

%==================== RIGHT: ASYMGENESIS ====================
\begin{scope}[xshift=13cm]

% ---- ORIGINAL ASYMGENESIS CODE (UNCHANGED) ----
\begin{scope}[scale=1.2]

\draw[domain=-1.185:1.185, smooth, thick]
  plot (\x, {-2*(\x)^2 + 2});

\draw[->,thick] (3.5,-0.5) -- (3.5,3.5) node[right] {$T$};
\draw[thick] (3.4,2) -- (3.6,2);
\node[left] at (3.4,2) {$T_{\mathrm{eq}}$};

\node (Lagrangian) at (0,2.5)
 {$\displaystyle \mathcal{L} \supset \frac{\mathcal{O}_C\mathcal{O}_{X}}{\Lambda^n}$};
\node [left=0.25cm of Lagrangian] {Charge transfer};

\node (SM Operator) at (-2,1)
 {$\displaystyle \mathcal{O}_C = \Bar{L}He,\dots$};
\node (SM text) [below=0cm of SM Operator] {SM sector};
\node [below=0cm of SM text] {$\mu_C \neq 0$};

\node (DM Operator) at (2,1)
 {$\displaystyle \mathcal{O}_X = X^p,\dots$};
\node (DM text) [below=0cm of DM Operator] {DM sector};
\node [below=0cm of DM text] {$\mu_X \neq 0$};

\draw[
  -{Latex[length=4mm, width=2mm]},
  line width=1.2pt
] (-3.5,1) -- (-4.5,1)
node[midway, above=0.1cm] {RHNs}
node (WILG) [midway, below=0.1cm] {WILG};

\node[draw, below=0cm of WILG, xshift=-0.3cm] {$B\!-\!L$ violation};

\node at (-5.5,1) {$\mu_{B-L} \neq 0$};

\end{scope}

% ---- Square bracket + label (Asymgenesis) ----
\draw[line width=0.8pt] (-7.6,-1.0) -- (-7.6,-0.5);
\draw[line width=0.8pt] ( 4.2,-1.0) -- ( 4.2,-0.5);
\draw[line width=0.8pt] (-7.6,-1.0) -- ( 4.2,-1.0);
\node at (-1.3,-2.0) {{\large\bfseries Asymgenesis}};

\end{scope}

\end{tikzpicture}
\caption{Left: Schematic of standard ADM models with charge transfer mediated by higher-dimensional operator(s) (adapted from Ref.~\cite{Zurek:2013wia}). Right: A schematic of charge transfer in Asymgenesis.}
\label{fig:ADM_Asymgenesis}
\end{figure*}

However, as will be shown in this work, the restrictions on the higher-dimensional interaction transferring asymmetry between the two sectors can be significantly relaxed in the context of the WILG paradigm. Rather than violating $B-L$, it is sufficient if the interaction takes the form 
\begin{align}
    \label{eq:operator}
    \mathcal{O}_{C}^{\rm{(SM)}}\mathcal{O}_X^{(D)}/\Tilde{\Lambda}^n+\text{h.c.}\,,
\end{align}
where $\mathcal{O}_C^{\rm{(SM)}}$ violates at least one of the conserved charges $C$ used in WILG\footnote{In many cases, $\mathcal{O}_C^{\rm{(SM)}}$ needs not even violate $C$, as explained in the Appendix.} and $\tilde{\Lambda}$ is a heavy mass scale. In general, $\mathcal{O}_{C}^{\rm{(SM)}}$ needs not be renormalizable or gauge invariant or Lorentz invariant as long as invariance is restored for the combination $\mathcal{O}_{C}^{\rm{(SM)}}\mathcal{O}_X^{(D)}$. However, in this letter, we will consider the simplest case where it is a Lorentz scalar, renormalizable, and gauge invariant, and thereby reduces to a SM Yukawa-type operator. To the best of our knowledge, this possibility has not been considered in the literature so far. In Appendix~\ref{Appendix}, we show that by relaxing these requirements one recovers a broader landscape of ADM constructions based on charge transfer via higher-dimensional operators, and we illustrate how the framework introduced in this letter can be applied to generalize established charge-transfer mechanisms. %In general, $\mathcal{O}_C^{\rm{(SM)}}$ can be any gauge-invariant operator, but for concreteness, we will consider the case where it is renormalizable, and reduces to a SM Yukawa-type operator. 
We will also consider the simplest realization possible for the dark operator $\mathcal{O}^{(D)}_X=X^p$, where $X$ is a complex scalar asymmetric DM candidate, and $\mathcal{O}_{X}^{(D)}$ violates "$X$-number" charge by $p$ units. In scenarios with $p>2$, potentially dangerous $X-\bar{X}$ oscillations~\cite{Buckley:2011ye,Cirelli:2011ac,Tulin:2012re} are autmatically avoided if $X$ is charged under a dark $\mathbb{Z}_p$ symmetry. Throughout, we assume that $X$ belongs to a dark sector that allows for sufficiently efficient annihilation of the symmetric abundance of $X$ such that the asymmetric component provides the dominant contribution to the DM density, and that the charge-transfer interaction in Eq.~\eqref{eq:operator} provides the dominant source of $X$-number violation. 

The framework introduced here, denoted \textit{Asymgenesis}, represents a unification of the conventional ADM scenario — formulated via higher-dimensional operators — with the type-I seesaw mechanism, establishing a common framework for neutrino mass generation and asymmetry transfer between sectors. The role of RHNs in Asymgenesis is the same as in WILG, namely to act on a nontrivial chemical background that features charge asymmetries in a $B-L$ violating manner. In contrast to standard ADM mechanisms, Asymgenesis allows for a separation between the scale of charge transfer between the dark and the visible sectors and $B-L$ violation. Indeed, Asymgenesis includes the intriguing possibility that the charge-transfer mechanism need not violate $B-L$. This possibility was also realized in Higgsogenesis~\cite{Servant:2013uwa} where a nonzero Higgs chemical potential -- induced by either a primordial $B-L$ asymmetry or a primordial dark charge -- mediates asymmetries between visible and dark sectors. Meanwhile, in Asymgenesis a primordial abundance of \textit{any} of the (global) SM charges that are approximately conserved at high temperatures suffices. In this respect, Asymgenesis provides a flexible charge-transfer framework with respect to the nature of the primordial asymmetry. The right part of Fig.~\ref{fig:ADM_Asymgenesis} presents a schematic summary of charge transfer in Asymgenesis, with further discussion on its relation to other ADM frameworks provided in Appendix~\ref{Appendix}. Finally, it is important to note that Asymgenesis serves solely as a framework for charge transfer between the visible and dark sector(s) in ADM models. Accordingly, any model implementation of Asymgenesis must comply with the usual phenomenological constraints on ADM, including those arising from the annihilation of symmetric dark matter components and from washout processes induced by particle–antiparticle oscillations in the dark sector. 

The remainder of the paper is organized as follows. We start by illustrating a simple prototype of the Asymgenesis framework before we generalize the description to arbitrary initial conditions and temperature regimes. First, we present results that are valid if the charge-transfer interaction induced by the operator in Eq.~\eqref{eq:operator} was in thermal equilibrium at high temperatures. We then go beyond the assumption of thermal equilibrium by deriving and solving Boltzmann equations that describe charge transfer from the SM to the DM sector and vice versa.

\begin{figure*}[t]
\centering
\begin{tikzpicture}[
  scale=0.8,
  transform shape,
  >=stealth,
  every node/.style={font=\normalsize}
]%[scale=0.8, transform shape, >=stealth]

%==================== LEFT: Asymgenesis I ====================
\begin{scope}[]

% ---- Figure ----
\begin{scope}[scale=0.9]

\draw[->] (5.5,0) -- (-6,0) node[below] {$T$};
\draw (5,0.08) -- (5,-0.08) node[below=4pt] {$T_{\rm EW}$};
\draw (1.5,0.08) -- (1.5,-0.08) node[below=4pt] {$T_{C}$};
\draw [dashed] (1.5,3) -- (1.5,-0.08);
\draw (0,0.08) -- (0,-0.08) node[below=4pt, xshift=+0.2cm] {$T_{B-L}$};
\draw (-0.5,0.08) -- (-0.5,-0.08) node[below=4pt, xshift=-0.2cm] {$M_{N_1}$};
\draw (-3,0.08) -- (-3,-0.08) node[below=4pt] {$T_{\rm eq}$};
\draw [dashed] (-3,3) -- (-3,-0.08);
\draw (-5,0.08) -- (-5,-0.08) node[below=4pt] {$T_{\rm CG}$};

% Box open on one side
\draw[thick] (2,3.75) -- (5,3.75) -- (5,4.25) -- (2,4.25);
% Dotted continuation of the top and bottom edges
\draw[thick, densely dotted] (2,3.75) -- ++(-0.5,0);
\draw[thick, densely dotted] (2,4.25)  -- ++(-0.5,0);
\node at (3,4) {Sphalerons ($B\!+\!L$ violation)};
% Optional ellipsis marker
% \node at (0-0.75,0.5*0.5) {$\cdots$};

% Box open on one side
\draw[thick] (3.5,3) -- (1.5,3) -- (1.5,3.5) -- (3.5,3.5);
% Dotted continuation of the top and bottom edges
\draw[thick, densely dotted] (3.5,3) -- ++(0.5,0);
\draw[thick, densely dotted] (3.5,3.5)  -- ++(0.5,0);
\node at (3.25,3.25) {Yukawa ($C$ violation)};
% Optional ellipsis marker
% \node at (0-0.75,0.5*0.5) {$\cdots$};

% Box open on one side
\draw[thick] (-3,3.75) -- (0,3.75) -- (0,4.25) -- (-3,4.25);
% Dotted continuation of the top and bottom edges
\draw[thick, densely dotted] (-3,3.75) -- ++(-0.5,0);
\draw[thick, densely dotted] (-3,4.25)  -- ++(-0.5,0);
\node at (-1.75,4) {RHNs ($B\!-\!L$ violation)};
% Optional ellipsis marker
% \node at (0-0.75,0.5*0.5) {$\cdots$};

% Box open on one side
\draw[thick] (-5,3) -- (-3,3) -- (-3,3.5) -- (-5,3.5);
% Dotted continuation of the top and bottom edges
\draw[thick, densely dotted] (-5,3) -- ++(-0.5,0);
\draw[thick, densely dotted] (-5,3.5)  -- ++(-0.5,0);
\node at (-4.75,3.25) {Portal ($XC$ violation)};
% Optional ellipsis marker
% \node at (0-0.75,0.5*0.5) {$\cdots$};

\node at (4.5,1.75) {$\mu_C = 0$};
\node at (4.5,1.25) {$\mu_X \neq 0$};
\node at (4.5,0.75) {$\mu_{B-L} \neq 0$};

\node at (0,1.75) {$\mu_C \neq 0$};
\node at (0,1.25) {$\mu_X \neq 0$};
\node at (0,0.75) {$\mu_{B-L} \neq 0$};

\node at (-2.25,1.75) {$\mu_C \neq 0$};
\node at (-2.25,1.25) {$\mu_X \neq 0$};
%\node at (-2.25,0.75) {$\mu_{B-L} = 0$};

\node at (-4,1.5) {$\mu_{XC} \neq 0$};
%\node at (-4,0.75) {$\mu_{B-L} = 0$};

\node at (-5.75,1.5) {$\mu_{XC} = 0$};
\node at (-5.75,0.75) {$\mu_{B-L} = 0$};

\end{scope}

% ---- Square bracket + label (ADM) ----
% \draw[line width=0.8pt] (-4.6,-1.0) -- (-4.6,-0.5);
% \draw[line width=0.8pt] ( 4.2,-1.0) -- ( 4.2,-0.5);
% \draw[line width=0.8pt] (-4.6,-1.0) -- ( 4.2,-1.0);
\node at (0,-1.5) {\large\bfseries $T_{\rm{eq}}>T_{B-L}$};

\end{scope}

\draw[thick] (-6,-1) -- (-6,4);
\draw[thick] (5,-1) -- (5,4);
\draw[thick] (16,-1) -- (16,4);

%==================== RIGHT: ASYMGENESIS II ====================
\begin{scope}[xshift=11cm]

% ---- Figure ----
\begin{scope}[scale=0.9]

\draw[->] (5.5,0) -- (-6,0) node[below] {$T$};
\draw (5,0.08) -- (5,-0.08) node[below=4pt] {$T_{\rm EW}$};
\draw (1.5,0.08) -- (1.5,-0.08) node[below=4pt] {$T_{C}$};
\draw [dashed] (1.5,3) -- (1.5,-0.08);
\draw (0,0.08) -- (0,-0.08) node[below=4pt] {$T_{\rm eq}$};
\draw [dashed] (0,3) -- (0,-0.08);
\draw (-3,0.08) -- (-3,-0.08) node[below=4pt, xshift=+0.2cm] {$T_{B-L}$};
\draw (-3.5,0.08) -- (-3.5,-0.08) node[below=4pt, xshift=-0.2cm] {$M_{N_1}$};
\draw (-5,0.08) -- (-5,-0.08) node[below=4pt] {$T_{\rm CG}$};

% Box open on one side
\draw[thick] (2,3.75) -- (5,3.75) -- (5,4.25) -- (2,4.25);
% Dotted continuation of the top and bottom edges
\draw[thick, densely dotted] (2,3.75) -- ++(-0.5,0);
\draw[thick, densely dotted] (2,4.25)  -- ++(-0.5,0);
\node at (3,4) {Sphalerons ($B\!+\!L$ violation)};
% Optional ellipsis marker
% \node at (0-0.75,0.5*0.5) {$\cdots$};

% Box open on one side
\draw[thick] (3.5,3) -- (1.5,3) -- (1.5,3.5) -- (3.5,3.5);
% Dotted continuation of the top and bottom edges
\draw[thick, densely dotted] (3.5,3) -- ++(0.5,0);
\draw[thick, densely dotted] (3.5,3.5)  -- ++(0.5,0);
\node at (3.25,3.25) {Yukawa ($C$ violation)};
% Optional ellipsis marker
% \node at (0-0.75,0.5*0.5) {$\cdots$};

% Box open on one side
\draw[thick] (-6,3.75) -- (-3,3.75) -- (-3,4.25) -- (-6,4.25);
% Dotted continuation of the top and bottom edges
\draw[thick, densely dotted] (-6,3.75) -- ++(-0.5,0);
\draw[thick, densely dotted] (-6,4.25)  -- ++(-0.5,0);
\node at (-4.75,4) {RHNs ($B\!-\!L$ violation)};
% Optional ellipsis marker
% \node at (0-0.75,0.5*0.5) {$\cdots$};

% Box open on one side
\draw[thick] (-2,3) -- (0,3) -- (0,3.5) -- (-2,3.5);
% Dotted continuation of the top and bottom edges
\draw[thick, densely dotted] (-2,3) -- ++(-0.5,0);
\draw[thick, densely dotted] (-2,3.5)  -- ++(-0.5,0);
\node at (-1.75,3.25) {Portal ($XC$ violation)};
% Optional ellipsis marker
% \node at (0-0.75,0.5*0.5) {$\cdots$};

\node at (4.5,1.75) {$\mu_C = 0$};
\node at (4.5,1.25) {$\mu_X \neq 0$};
\node at (4.5,0.75) {$\mu_{B-L} \neq 0$};

\node at (0.75,1.75) {$\mu_C \neq 0$};
\node at (0.75,1.25) {$\mu_X \neq 0$};
\node at (0.75,0.75) {$\mu_{B-L} \neq 0$};

\node at (-3,1.5) {$\mu_{XC} \neq 0$};
\node at (-3,0.75) {$\mu_{B-L} \neq 0$};

%\node at (-4,1.5) {$\mu_{XC} \neq 0$};
%\node at (-4,0.75) {$\mu_{B-L} = 0$};

\node at (-5.75,1.5) {$\mu_{XC} = 0$};
\node at (-5.75,0.75) {$\mu_{B-L} = 0$};

\end{scope}

% ---- Square bracket + label (Asymgenesis) ----
% \draw[line width=0.8pt] (-7.6,-1.0) -- (-7.6,-0.5);
% \draw[line width=0.8pt] ( 4.2,-1.0) -- ( 4.2,-0.5);
% \draw[line width=0.8pt] (-7.6,-1.0) -- ( 4.2,-1.0);
\node at (0,-1.5) {{\large\bfseries $T_{B-L}>T_{\rm{eq}}$}};

\end{scope}

\end{tikzpicture}
\caption{Schematic evolution of chemical potentials in Asymgenesis for $T_{\rm{eq}}>T_{B-L}$ (left) and $T_{B-L}>T_{\rm{eq}}$ (right). Here, $T_{\rm{CG}},\,T_{B-L},\,T_{\rm{eq}},\,T_C,$ and $T_{\rm{EW}}$ denote the temperature scales associated with chargegenesis, RHN interactions becoming inefficient, decoupling of charge transfer between the dark and visible sectors, $C$-violating interactions becoming efficient, and the electroweak phase transition, respectively.}
\label{fig:Asymgenesis_versions}
\end{figure*}

\noindent\textbf{Warmup}\,---\, We initiate our analysis by examining the transfer of charge between the visible and the dark sector. The charge asymmetry $q_i$ of a SM or DM particle species $i$ is defined as, $q_i=n_i-n_{\bar{i}}=g_i\mu_iT^2/6$,\footnote{This formula is valid in the limit $T\gg m_i$, and is justified by the fact that all relevant charge-generation and charge-transfer processes must occur at $T\gtrsim 100$ TeV in the scenario to be considered and that $m_X\lesssim 10$ TeV from perturbative unitarity.} where $n_i$ $(n_{\bar{i}})$ denotes the (anti)particle number density of particle $i$ with chemical potential $\mu_i$ and multiplicity $g_i$.
At high temperatures, the interaction described in Eq.~\eqref{eq:operator} may establish thermal equilibrium between the two sectors, 
relating their chemical potentials $\mu_i$~\cite{Harvey:1990qw}. Further, all SM and DM chemical potentials can be expressed as linear combinations of a set of linearly independent conserved charges $q_C = \bar{\mu}_C T^2 / 6$, e.g. by employing the linear-algebra formalism in~\cite{Domcke:2020quw} (see also~\cite{Antaramian:1993nt,Fong:2015vna} for some related earlier works). Here, $\bar{\mu}_C$ is introduced for notional convenience: $\bar{\mu}_C$ and $q_C$ are not conjugate to each other in the usual thermodynamical sense, but both describe the conserved charge. In other words, $\bar{\mu}_C$ corresponds to $q_C$, written in units of a chemical potential.
To illustrate the concept of Asymgenesis, we consider a specific example where $\mathcal{O}_{C}^{\rm{(SM)}}=\overline{L_e}He_R$, with $L_e$, $H$, and $e_R$ representing the first-generation lepton doublet, the SM Higgs doublet, and the right-handed electron, respectively. If the higher-dimensional operator is in thermal equilibrium, the following equilibrium condition relating the SM and the dark sector holds,
\begin{align}
    p\cdot\mu_X=-\mu_H-\mu_e+\mu_{L_e}.
\end{align}
While the operator $\mathcal{O}_{C}^{\rm{(SM)}}=\overline{L_e}He_R$ violates right-handed electron number, the interaction Eq.~\eqref{eq:operator} conserves a combination of right-handed electron number and dark number. Consequently, the conservation of the right-handed electron charge, $q_e\sim \mu_e/T$, at high temperatures~\cite{Campbell:1992jd,Cline:1993vv,Cline:1993bd,Bodeker:2019ajh} is modified as follows
\begin{align}
    \frac{\mu_e-(2/p)\cdot\mu_X}{T}\equiv \frac{\Bar{\mu}_{Xe}}{T}=\text{constant},
\end{align}
where we used $g_X=2$, which we employ throughout this paper.
As an example, if the charge-transfer interaction decouples at temperatures of a few $100$ TeV, then the ADM chemical potential can be written as
\begin{align}
    \label{muX2}
    \frac{1422+481p^2}{p}\mu_X=-711\Bar{\mu}_{Xe}-185\bar{\mu}_{\Delta_e}+52\left(\bar{\mu}_{\Delta_\mu}+\bar{\mu}_{\Delta_\tau}\right).
\end{align}
The ADM abundance would in this case be determined by two factors: (i) the primordial asymmetries in $q_e$ and $q_X$, and (ii) the values of the flavored $B-L$ asymmetries, $\Delta_\alpha$, at the time of charge-transfer decoupling. 
If the lightest RHN, $N_1$, has a mass around $M_N\sim10^{2-3}$ TeV and decouples while the portal interaction is still efficient, then WILG in the strong wash-in regime would lead to
\begin{align}
    \begin{pmatrix}
        \bar{\mu}_{\Delta_e}\\
        \bar{\mu}_{\Delta_\mu}\\
        \bar{\mu}_{\Delta_\tau}
    \end{pmatrix}=\begin{pmatrix}
        -25\\
         8\\
        8
    \end{pmatrix} \frac{p^2}{\left(66+30p^2\right)}\Bar{\mu}_{Xe},
\end{align}
and both the ADM abundance and the BAU would be fixed entirely by the primordial asymmetry in $q_{Xe}$. The latter could be generated through a CG mechanism that generates nonzero $q_e$ and/or nonzero $q_X$. At temperatures above $10^6$ GeV, additional SM charges are approximately conserved, and can be used as part of Asymgenesis.

\begin{table*}[]%[t]
\caption{The numerical coefficients $x_C$ (\textcolor{Bittersweet}{brown}) and $y_C$ (\textcolor{ForestGreen}{green}), associated with the conserved charges $\bar{\mu}_C=q_C6/T^2$ as defined in Eq.~\eqref{eq:qBLAndqXEq} for the scenario with $\mathcal{O}_{C}^{\rm{(SM)}}=\overline{L_e}He_R$ and $\mathcal{O}_{X}^{(D)}=X^p$ in different temperature regimes. 
The first column shows $T_{\rm{dec}}$, which is identified with $T_{B-L}$ $(T_{\rm{eq}})$ for $\bar{\mu}_{B-L}^{\rm eq}$ $\left(\mu_{X}^{\rm eq}\right)$, see text below Eq.~\eqref{eq:qBLAndqXEq}, and (i)-(v) indicate the following ranges for $T_{\rm{dec}}$ [GeV]: $\qty(10^5, 10^6)$, $\qty(10^6, 10^9)$, $\qty(10^9, 10^{11-12})$, $\qty(10^{11-12}, 10^{13})$, $\qty(10^{13}, 10^{15})$, respectively.
The flavor index $\alpha$ in the second column lists the uncorrelated charged-lepton flavors that interact with $N_1$ in the relevant temperature range, assuming the SM–DM portal is active (decoupled) in that range. In the latter case, the results for $x_C$ reduce to those in standard WILG, c.f. Table II in~\cite{Domcke:2020quw}, which are included in parentheses in column $3$ below. 
The \xmark~symbol marks the absence of the corresponding $\bar{\mu}_C$ due to an efficient SM interaction.
The coefficients $P$ and $P_{e,\tau}$ encode the primordial $q_{e,\mu,\tau}$ asymmetries’ flavor composition relative to the $N_1$ washout direction, as detailed in~\cite{Domcke:2020quw}. See text for further details. 
}
\begin{tabular}{|c|c|ccc|c|c}\hline
	 $T_{\rm{dec}}$ & Index $\alpha$ & \makecell{$\bar{\mu}_{Xe}$ ($\bar{\mu}_e$) \\ $\bar{\mu}_{2 B_1 - B_2 - B_3}$\\ $\bar{\mu}_{u-d}$ \\$\bar{\mu}_{d-s}$} & \makecell{$\bar{\mu}_{B_1 - B_2}$ \\  $\bar{\mu}_\mu$ \\  $\bar{\mu}_{u-c}$\\  $\bar{\mu}_\tau$}  & \makecell{ $\bar{\mu}_{d-b}$\\$\bar{\mu}_B$\\ $\bar{\mu}_u$\\   $\bar{\mu}_{\Delta_\perp}$}   & $\bar{\mu}_{\Delta_\alpha}$  \\[.2em]
	\hline
	(i) & $e,\mu,\tau$ $(e,\mu,\tau)$ & \makecell{\color{Bittersweet}{$\frac{-3p^2}{22+10p^2}$} {\color{Bittersweet}{$(\frac{-3}{10})$},\color{ForestGreen}{$\frac{-711}{1422+481p^2}$}}\\ \color{Bittersweet}{\xmark} \color{Bittersweet}{(\xmark)},\color{ForestGreen}{\xmark} \\ \color{Bittersweet}{\xmark} \color{Bittersweet}{(\xmark)},\color{ForestGreen}{\xmark}\\ \color{Bittersweet}{\xmark} \color{Bittersweet}{(\xmark)},\color{ForestGreen}{\xmark}}& \makecell{ \color{Bittersweet}{\xmark} \color{Bittersweet}{(\xmark)},\color{ForestGreen}{\xmark} \\ \color{Bittersweet}{\xmark} \color{Bittersweet}{(\xmark)},\color{ForestGreen}{\xmark} \\ \color{Bittersweet}{\xmark} \color{Bittersweet}{(\xmark)},\color{ForestGreen}{\xmark}\\ \color{Bittersweet}{\xmark} \color{Bittersweet}{(\xmark)},\color{ForestGreen}{\xmark}}& \makecell{ \color{Bittersweet}{\xmark} \color{Bittersweet}{(\xmark)},\color{ForestGreen}{\xmark}\\ \color{Bittersweet}{\xmark} \color{Bittersweet}{(\xmark)},\color{ForestGreen}{\xmark}\\ \color{Bittersweet}{\xmark} \color{Bittersweet}{(\xmark)},\color{ForestGreen}{\xmark} \\ \color{Bittersweet}{\xmark} \color{Bittersweet}{(\xmark)},\color{ForestGreen}{\xmark}}  & \color{ForestGreen}{$\frac{[-185,\,52,\,52]}{1422+481p^2}$}\\ \hline
	 (ii) & $e,\mu,\tau$ $(e,\mu,\tau)$ & \makecell{\color{Bittersweet}{$\frac{-3p^2}{38+17p^2}$} \color{Bittersweet}{$\left(\frac{-3}{17}\right)$},\color{ForestGreen}{$\frac{-1071}{2142+716p^2}$}\\ \color{Bittersweet}{$0$} \color{Bittersweet}{$(0)$}, \color{ForestGreen}{$0$} \\ \color{Bittersweet}{$\frac{-16-7p^2}{38+17p^2}$} \color{Bittersweet}{$\left(\frac{-7}{17}\right)$}, \color{ForestGreen}{$\frac{45}{2142+716p^2}$} \\ \color{Bittersweet}{\xmark} \color{Bittersweet}{(\xmark)},\color{ForestGreen}{\xmark}}& \makecell{ \color{Bittersweet}{\xmark} \color{Bittersweet}{(\xmark)},\color{ForestGreen}{\xmark} \\ \color{Bittersweet}{\xmark} \color{Bittersweet}{(\xmark)},\color{ForestGreen}{\xmark} \\ \color{Bittersweet}{\xmark} \color{Bittersweet}{(\xmark)},\color{ForestGreen}{\xmark}\\ \color{Bittersweet}{\xmark} \color{Bittersweet}{(\xmark)},\color{ForestGreen}{\xmark}}& \makecell{ \color{Bittersweet}{\xmark} \color{Bittersweet}{(\xmark)},\color{ForestGreen}{\xmark} \\ \color{Bittersweet}{\xmark} \color{Bittersweet}{(\xmark)},\color{ForestGreen}{\xmark}\\ \color{Bittersweet}{\xmark} \color{Bittersweet}{(\xmark)},\color{ForestGreen}{\xmark} \\ \color{Bittersweet}{\xmark} \color{Bittersweet}{(\xmark)},\color{ForestGreen}{\xmark}}  & \color{ForestGreen}{$\frac{[-265,\,92,\,92]}{2142+716p^2}$}\\ \hline
	(iii) & $e,\mu,\tau$ $(\parallel_\tau,\tau)$  
	& \makecell{\color{Bittersweet}{$\frac{-3p^2}{62+27p^2}$} \color{Bittersweet}{$\left(\frac{142-225P_\tau}{247}\right)$},\color{ForestGreen}{$\frac{-1779}{3558+1178p^2}$}\\ \color{Bittersweet}{$0$} \color{Bittersweet}{$(0)$},\color{ForestGreen}{$0$} \\ \color{Bittersweet}{$\frac{-6(7+3p^2)}{62+27p^2}$} \color{Bittersweet}{$\left(\frac{-123}{247}\right)$},\color{ForestGreen}{$\frac{135}{3558+1178p^2}$} \\ \color{Bittersweet}{$\frac{-4(7+3p^2)}{62+27p^2}$}\color{Bittersweet}{$\left(\frac{-82}{247}\right)$},\color{ForestGreen}{$\frac{45}{1779+589p^2}$}} & \makecell{ \color{Bittersweet}{$\frac{21+9p^2}{62+27p^2}$} \color{Bittersweet}{$\left(\frac{123}{494}\right)$},\color{ForestGreen}{$\frac{-135}{7116+2356p^2}$} \\ \color{Bittersweet}{$\frac{-3(2+p^2)}{62+27p^2}$} \color{Bittersweet}{$\left(\frac{142-225P_\tau}{247}\right)$},\color{ForestGreen}{$\frac{-6}{1779+589p^2}$}\\ \color{Bittersweet}{\xmark} \color{Bittersweet}{(\xmark)},\color{ForestGreen}{\xmark}\\ \color{Bittersweet}{\xmark}\color{Bittersweet}{(\xmark)},\color{ForestGreen}{\xmark}}& \makecell{ \color{Bittersweet}{\xmark}\color{Bittersweet}{(\xmark)},\color{ForestGreen}{\xmark} \\ \color{Bittersweet}{\xmark}\color{Bittersweet}{(\xmark)},\color{ForestGreen}{\xmark}\\ \color{Bittersweet}{\xmark}\color{Bittersweet}{(\xmark)},\color{ForestGreen}{\xmark} \\ \color{Bittersweet}{\xmark} \color{Bittersweet}{$\left(\frac{225}{247}\right)$},\color{ForestGreen}{\xmark}} & \color{ForestGreen}{$\frac{[-421,\,168,\,172]}{3558+1178p^2}$}\\ \hline
	 (iv) & $e,\parallel_e$ $(\parallel)$  & \makecell{\color{Bittersweet}{$\frac{-9p^2}{102+37p^2}$} \color{Bittersweet}{$\left(\frac{-23P+7}{30}\right)$},\color{ForestGreen}{$\frac{-38}{76+23p^2}$}\\ \color{Bittersweet}{$\frac{12(3+p^2)}{102+37p^2}$} \color{Bittersweet}{$\left(\frac{1}{5}\right)$},\color{ForestGreen}{$\frac{-3}{76+23p^2}$} \\ \color{Bittersweet}{$\frac{-36(3+p^2)}{102+37p^2}$} \color{Bittersweet}{$\left(\frac{-3}{5}\right)$},\color{ForestGreen}{$\frac{9}{76+23p^2}$} \\ \color{Bittersweet}{$\frac{-10(3+p^2)}{102+37p^2}$} \color{Bittersweet}{$\left(\frac{-1}{6}\right)$},\color{ForestGreen}{$\frac{5}{152+46p^2}$}}& \makecell{ \color{Bittersweet}{$\frac{-18(3+p^2)}{102+37p^2}$} \color{Bittersweet}{$\left(\frac{-3}{10}\right)$},\color{ForestGreen}{$\frac{9}{152+46p^2}$} \\ \color{Bittersweet}{$\frac{42+p^2(14-23P_e)-60P_e}{102+37p^2}$} \color{Bittersweet}{$\left(\frac{-23P+7}{30}\right)$},\color{ForestGreen}{$\frac{-7}{152+46p^2}$} \\\color{Bittersweet}{$\frac{18(3+p^2)}{102+37p^2}$} \color{Bittersweet}{$\left(\frac{3}{10}\right)$},\color{ForestGreen}{$\frac{-9}{152+46p^2}$}\\ \color{Bittersweet}{$\frac{42+p^2(14-23P_e)-60P_e}{102+37p^2}$} \color{Bittersweet}{$\left(\frac{-23P+7}{30}\right)$},\color{ForestGreen}{$\frac{-7}{152+46p^2}$}}& \makecell{ \color{Bittersweet}{$\frac{-16(3+p^2)}{102+37p^2}$} \color{Bittersweet}{$\left(\frac{-4}{15}\right)$},\color{ForestGreen}{$\frac{4}{76+23p^2}$} \\ \color{Bittersweet}{$\frac{120+46p^2}{306+111p^2}$} \color{Bittersweet}{$\left(\frac{23}{90}\right)$},\color{ForestGreen}{$\frac{23/6}{76+23p^2}$}\\ \color{Bittersweet}{\xmark} \color{Bittersweet}{(\xmark)},\color{ForestGreen}{\xmark} \\ \color{Bittersweet}{$\frac{60+23p^2}{102+37p^2}$} \color{Bittersweet}{$\left(\frac{23}{30}\right)$},\color{ForestGreen}{$\frac{7}{152+46p^2}$}} & \color{ForestGreen}{$\frac{[-16,7]}{152+46p^2}$}\\ \hline
	 (v) & $e,\parallel_e$ $(\parallel)$ &\makecell{\color{Bittersweet}{$\frac{-p^2}{14+5p^2}$} \color{Bittersweet}{$\left(\frac{-3P+1}{4}\right)$},\color{ForestGreen}{$\frac{-5}{10+3p^2}$}\\\color{Bittersweet}{$\frac{4(3+p^2)}{42+15p^2}$} \color{Bittersweet}{$\left(\frac{1}{6}\right)$}, \color{ForestGreen}{$\frac{-1}{30+9p^2}$}\\\color{Bittersweet}{$\frac{-20(3+p^2)}{42+15p^2}$}\color{Bittersweet}{$\left(\frac{-5}{6}\right)$},\color{ForestGreen}{$\frac{5}{30+9p^2}$} \\ \color{Bittersweet}{$\frac{-2(3+p^2)}{14+5p^2}$} \color{Bittersweet}{$\left(\frac{-1}{4}\right)$}, \color{ForestGreen}{$\frac{1}{20+6p^2}$}}&\makecell{ \color{Bittersweet}{$\frac{-2(3+p^2)}{14+5p^2}$} \color{Bittersweet}{$\left(\frac{-1}{4}\right)$},\color{ForestGreen}{$\frac{1}{20+6p^2}$} \\ \color{Bittersweet}{$\frac{6+p^2(2-3P_e)-8P_e}{14+5p^2}$} \color{Bittersweet}{$\left(\frac{-3P+1}{4}\right)$},\color{ForestGreen}{$\frac{-1}{20+6p^2}$} \\ \color{Bittersweet}{$\frac{2(3+p^2)}{14+5p^2}$} \color{Bittersweet}{$\left(\frac{1}{4}\right)$},\color{ForestGreen}{$\frac{-1}{20+6p^2}$} \\ \color{Bittersweet}{$\frac{6+p^2(2-3P_e)-8P_e}{14+5p^2}$} \color{Bittersweet}{$\left(\frac{-3P+1}{4}\right)$}, \color{ForestGreen}{$\frac{-1}{20+6p^2}$}}&\makecell{\color{Bittersweet}{$\frac{-8(3+p^2)}{42+15p^2}$} \color{Bittersweet}{$\left(\frac{-1}{3}\right)$},\color{ForestGreen}{$\frac{2}{30+9p^2}$}\\ \color{Bittersweet}{$\frac{10+4p^2}{42+15p^2}$} \color{Bittersweet}{$\left(\frac{1}{6}\right)$},\color{ForestGreen}{$\frac{2}{30+9p^2}$}\\ \color{Bittersweet}{$\frac{8(3+p^2)}{42+15p^2}$} \color{Bittersweet}{$\left(\frac{1}{3}\right)$}, \color{ForestGreen}{$\frac{-2}{30+9p^2}$} \\ \color{Bittersweet}{$\frac{8+3p^2}{14+5p^2}$} \color{Bittersweet}{$\left(\frac{3}{4}\right)$},\color{ForestGreen}{$\frac{1}{20+6p^2}$}} &\color{ForestGreen}{$\frac{[-2,\,1]}{20+6p^2}$}\\\hline
\end{tabular}
\label{tab:toolkit}
\end{table*}

\noindent\textbf{Asymgenesis}\,---\,
Consider a general scenario where primordial charges in both the visible and the dark sector can take arbitrary values and the charge-transfer interaction is in thermal equilibrium at high temperatures. Schematically, this corresponds to the regime above the barrier in the right panel of Fig.~\ref{fig:ADM_Asymgenesis}. The $B-L$ asymmetry in the strong wash-in regime and the ADM asymmetry can then, in full generality, be expressed as
\begin{align}
    \label{eq:qBLAndqXEq}
    \frac{q_{B-L}^{\rm{win}}}{s}=\sum_{C\neq \Delta_\alpha}x_C\frac{q_C}{s}\Bigg|_{T_{B-L}}, \quad \frac{q_X}{s}=2p\sum_{C}y_C\frac{q_C}{s}\Bigg|_{T_{\text{eq}}},%\frac{q_X}{s}=\frac{2}{p}\sum_{C}y_C\frac{q_C}{s}\Bigg|_{T_{\text{eq}}},
\end{align}
where $s$ denotes entropy density, and $T_{B-L}$ $(T_{\rm{eq}})$ denotes the temperature when RHN interactions (the charge-transfer interaction) become(s) inefficient. Here, $\Delta_\alpha=B/3-L_\alpha$ denote lepton-flavor asymmetries corresponding to the flavor combinations that are decohered at the relevant temperature scale.
At temperatures below $T_{\rm{eq}}$, corresponding to the regime below the top of the barrier in Fig.~\ref{fig:ADM_Asymgenesis}, both $q_e$ and $q_X$ become separately conserved. In particular, for $T_{B-L}<T_{\rm{eq}}$, $C=Xe$ can be replaced with $C=e$ in Eq.~\eqref{eq:qBLAndqXEq}, with $q_e$ given by
\begin{align}
    \frac{q_e}{s}=\frac{q_{Xe}}{s}+2\sum_{C}y_C\frac{q_C}{s}\Bigg|_{T_{\text{eq}}}.
\end{align}
In this case, the numerical values for $x_C$ reduce to the result from standard WILG~\cite{Domcke:2020quw}. Figure~\ref{fig:Asymgenesis_versions} provides an overview of the thermal history in Asymgenesis, highlighting the characteristic temperature regimes and the evolution of charge asymmetries. The left panel shows the scenario where $T_{B-L}>T_{\rm{eq}}$, while the right panel shows the opposite ordering $T_{\rm{eq}}>T_{B-L}$. Moreover, the values for $x_C$ and $y_C$ are shown for various temperature regimes in Table~\ref{tab:toolkit} for our example where $\mathcal{O}_{C}^{\rm{(SM)}}=\overline{L_e}He_R$\footnote{It is worth noting that if the second or third generation of leptons should enter $\mathcal{O}_{C}^{\rm{(SM)}}$ instead of the first generation as considered here, then the charge-transfer interaction must decouple above the equilibration temperatures of the $\mu$ or the $\tau$ Yukawa interactions $T_\mu=10^9$ GeV or $T_\tau=10^{12}$ GeV, respectively, to avoid large washout of the asymmetric DM component. A detailed account of asymmetry generation for different choices of $\mathcal{O}_{C}^{\rm{(SM)}}$ will be presented elsewhere~\cite{MojahedAndWeber}.}.
The coefficients without parentheses correspond to $T_{\rm{CG}}>T_{B-L}>T_{\rm{eq}}$, while the coefficients in parentheses correspond to $T_{\rm{CG}}>T_{\rm{eq}}>T_{B-L}$, where $T_{\rm{CG}}$ is the temperature scale associated with CG. 
The table shows that both the asymmetry in $X$ particles and the BAU are typically related to the primordial charge asymmetries generated during CG by an $\mathcal{O}(0.1-1)$ number. The analogous statement in the context of standard ADM models is that if the $B-L$ and $X$-violating higher-dimensional operator is in thermal equilibrium at high temperatures, then the BAU is related to the asymmetric component of DM by a $\mathcal{O}(1)$ number, barring potential washout processes. In conclusion, the framework considered so far predicts DM masses of $\mathcal{O}(1)-\mathcal{O}(10)$ GeV, which is similar to standard ADM models.

\noindent\textbf{Beyond thermal equilibrium: SM to DM}\,---\,Up to this point, we have analyzed the exchange of charges between the SM and the dark sector under the assumption that the charge-transfer interaction attained thermal equilibrium in the early universe. This requirement, however, is not essential. The same dynamics can be treated in an out-of-equilibrium framework using semiclassical Boltzmann equations. In this section, we relax the assumption of thermal equilibrium and consider a scenario in which the primordial charges in the dark sector are negligible. In the next section, we study the opposite limit, where the primordial charges in the SM sector are negligible. More generally, however, we emphasize that a CG mechanism can generate nonzero primordial charges in both the SM and the dark sector. 
To facilitate this description, we introduce the following quantities
\begin{align}
    \label{Ydef}
    Y_{\Delta i}&=\frac{q_i}{s},\quad Y_C=\frac{q_C}{s}, \quad z=\frac{M_N}{T}.
\end{align}
We find that the Boltzmann equation governing charge transfer from the SM plasma to the dark sector can be conveniently written as follows, 
\begin{align}\label{eq:result}
    \frac{dY_{\Delta X}}{dz}=\left[Y_{\Delta X}^{\rm eq}(z)-Y_{\Delta X}(z)\right]W(z),
\end{align}
where 
\begin{align}\label{eq:YDeltaXEq}
    Y_{\Delta X}^{\rm eq}(z)=2p\sum_{C}y_C(z)Y_C(z).
\end{align}
The washout term, $W(z)$, associated with the interactions induced by Eq.~\eqref{eq:operator} takes the following form,\footnote{The expression for $\Gamma^w$ is only valid in the regime where $T<\Lambda$, while for $T\gtrsim \Lambda$ it would be expected to take the following form $\Gamma^w=cT$, where $c$ is a constant.} 
\begin{align}\label{eq:wash_out_def}
    W=\frac{\Gamma^w}{zH}, \quad \Gamma^w=\frac{M_N^{2n+1}}{z^{2n+1}\Lambda^{2n}}, \quad H=\sqrt{\frac{\pi^2g_*}{90}}\frac{M_N^2}{z^2M_P},
\end{align}
assuming $X$ is relativistic in the relevant temperature regime. Here, $H$ is the Hubble parameter for a radiation-dominated universe, $g_*$ is the number of relativistic degrees of freedom, and $M_P$ denotes the reduced Planck mass. Finally, $\Lambda\sim\Tilde{\Lambda}$, up to a model-dependent numerical prefactor, where $\Tilde{\Lambda}$ is the effective-operator scale introduced in Eq.~\eqref{eq:operator}. The solution to Eq.~\eqref{eq:result} takes the following form
\begin{align}
    \label{AnalyticalPlanck}
Y_{\Delta X}(z) &= \frac{1}{\omega(z)} \left( Y_{\Delta X}(z_{\text{CG}})+ \int_{z_{\text{CG}}}^{z} \omega(\tilde{z}) W(\tilde{z}) Y_{\Delta X}^{\rm{eq}}(\tilde{z}) \, \mathrm{d}\tilde{z}  \right),
\end{align}
where
\begin{align}
    \label{mu(z)}
\omega(z) &= \exp\left(\int_{z_{\text{CG}}}^{z} W(\tilde{z}) \, \mathrm{d}\tilde{z}\right),
\end{align}
and $z_{\rm{CG}}=M_N/T_{\rm{CG}}$. As the electron-Yukawa interaction comes into equilibrium, $Y_{\Delta X}^{\rm{eq}}$ goes to zero. If, at that stage, the charge-transfer interaction characterized by $\Gamma^w$ remains efficient, Eqs.~\eqref{AnalyticalPlanck} and~\eqref{mu(z)} show that $Y_{\Delta X}$ is driven to zero, thereby washing out the 
$X$ asymmetry. Meanwhile, if the charge-transfer interactions are initially in thermal equilibrium and decouple above the electron Yukawa equilibration temperature, $T_e\approx 85$ TeV~\cite{Bodeker:2019ajh}, then the final ADM yield, $Y_{\Delta X}^{\rm{end}}$, is driven exponentially close to $Y_{\Delta X}^{\rm{eq}}$, as given by Eq.~\eqref{eq:YDeltaXEq} and Table~\ref{tab:toolkit}. The attraction of $Y_{\Delta X}^{\rm{end}}$ towards $Y_{\Delta X}^{\rm{eq}}$ driven by $W$, can be referred to as \textit{dark wash-in}, see also~\cite{Asadi:2025vli}.

\begin{figure}
    \centering
    \includegraphics[width=0.99\linewidth]{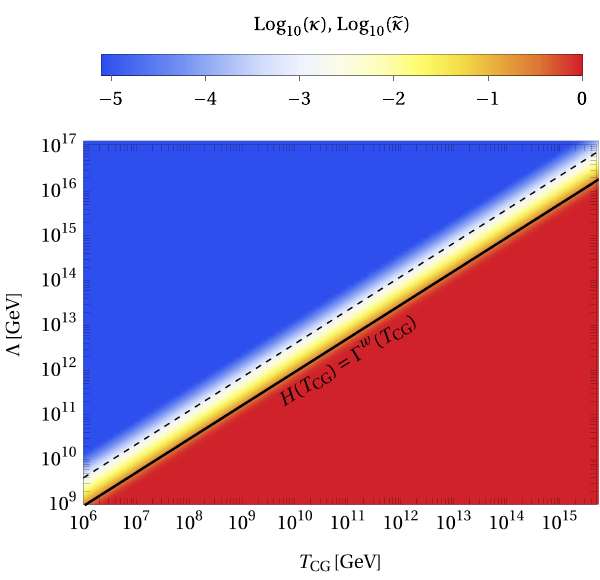}
    \caption{Numerical result for $\kappa$ $(\Tilde{\kappa})$, defined in Eq.~\eqref{YDeltaXFinal} (Eq.~\eqref{Yekappadef}) for charge transfer from the SM (DM) sector to the DM (SM) sector, in the $T_{\rm{CG}}-\Lambda$ plane. See text for details.} 
    \label{fig:parameter_space}
\end{figure}

Suppose instead that the effective-operator scale is so high that the charge-transfer interaction never enters thermal equilibrium. The charge transfer then becomes a UV freeze-in (FI) process~\cite{Giudice:2000ex,Hall:2009bx,Elahi:2014fsa}. In such a case, $Y_{\Delta X}^{\rm{end}}$ is to a good approximation, determined by the value of $Y_{\Delta X}^{\rm{eq}}$ right after CG, and the value of $\Lambda$ and $T_{\rm{CG}}$. To unify the UV FI regime and the thermal-equilibrium regime, we find it useful to express the final asymmetry in $\Delta X$ as
\begin{align}
    \label{YDeltaXFinal}
    Y_{\Delta X}^{\rm{end}}=\kappa(z')Y_{\Delta X}^{\rm{eq}}(z'),
\end{align}
where $z'=\text{Max}[z_{\rm{CG}},z_{\text{eq}}]$ and $\kappa$ parametrizes the charge-transfer efficiency.  
Numerical values for $\kappa$ for the case, $\mathcal{O}_{X}^{(D)}=X^2$, are shown in Fig.~\ref{fig:parameter_space}, where the horizontal and vertical axes show $T_{\rm{CG}}$ and $\Lambda$, respectively, and the black solid curve shows $H(z_{\rm{CG}})=\Gamma^w(z_{\rm{CG}})$. The latter marks the transition between the two regimes: The area below the curve corresponds to the thermal-equilibrium regime, $\Gamma^{w}(T_{\rm CG})/H(T_{\rm CG}) >1$, where $\kappa\simeq 1$ and the asymmetry is well described by Tab.~\ref{tab:toolkit}. The area above the curve shows the UV FI regime, $\Gamma^{w}(T_{\rm CG})/H(T_{\rm CG}) < 1$, and in the limit $\Gamma^{w}(T_{\rm CG})/H(T_{\rm CG}) \ll 1$, we have $\kappa \simeq \Gamma^{w}(T_{\rm CG})/H(T_{\rm CG})$. 
For charge transfer from the SM to the DM sector, the asymmetry in the dark sector is significantly smaller than the BAU in large portions of parameter space where Asymgenesis proceeds via UV FI. The black-dashed curve shows the contour corresponding to $\kappa=10^{-3}$, which would point to DM masses of $\mathcal{O}(1)$ TeV. This roughly represents an upper bound on ADM masses when the symmetric annihilation cross section is required to be large enough for the ADM component to be the dominant contribution to the DM relic density while respecting perturbative unitarity~\cite{Griest:1989wd,Cohen:2010kn,Petraki:2013wwa}. Interestingly, for $T_{\rm{CG}}=10^{14-15}$ GeV, it could be possible to account for the DM abundance even if the scale of NP associated with charge transfer is related to the scale of grand unification or possibly quantum-gravity effects, c.f.\ the upper-right corner of the figure.
In summary, the region below the black solid curve in Fig.~\ref{fig:parameter_space} corresponds to DM masses in the range of $\mathcal{O}(1)-\mathcal{O}(10)$ GeV, while the area below the black dashed curve corresponds to DM masses up to $\mathcal{O}(1)$ TeV. Above the black dashed curve, the symmetric component cannot annihilate efficiently due to the constraints imposed by perturbative unitarity. Finally, if the symmetric component cannot be annihilated efficiently, the dark matter mass must be smaller than the values quoted here to avoid overclosing the universe.

\noindent\textbf{Beyond thermal equilibrium: DM to SM}\,---\, Next, we consider the final possibility in which the initial conditions are set by a CG mechanism in the dark sector, and the effective operator scale is sufficiently high that the charge-transfer interaction never reaches thermal equilibrium. For the example where $\mathcal{O}_{C}^{\rm{(SM)}}=\overline{L_e}He_R$, we obtain a Boltzmann equation for the charge transfer to the SM sector of the following form, 
\begin{align}\label{eq:result2}
    \frac{dY_{\Delta e}}{dz}=\left[Y_{\Delta e}^{\rm{eq}}(z)-Y_{\Delta e}(z)\right]\widetilde{W}(z),
\end{align}
where
\begin{align}
    Y_{\Delta e}^{\rm{eq}}(z)=Y_{Xe}(z)+2\sum_{C}y_CY_C(z),
\end{align}
and $\widetilde{W}(z)$ takes the same form as in Eq.~\eqref{eq:wash_out_def} although the numerical value of the scale $\Lambda$ entering $\widetilde{W}$ might differ from that entering $W$ by an $\mathcal{O}(1)$ number. In an abuse of notation, we will denote both the heavy mass scale entering $W$ and $\widetilde{W}$ by $\Lambda$ in our results. With this, the solution to
Eq.~\eqref{eq:result2} reads
\begin{align}
    \label{AnalyticalPlanck2}
Y_{\Delta e}(z) &= \frac{1}{\tilde{\omega}(z)} \left( Y_{\Delta e}(z_{\text{CG}})+ \int_{z_{\text{CG}}}^{z} \tilde{\omega}(\tilde{z}) \widetilde{W}(\tilde{z}) Y_{\Delta e}^{\rm{eq}}(\tilde{z}) \, \mathrm{d}\tilde{z}  \right),
\end{align}
where
\begin{align}
    \label{mu(z)2}
\tilde{\omega}(z) &= \exp\left(\int_{z_{\text{CG}}}^{z}  \widetilde{W}(\tilde{z}) \, \mathrm{d}\tilde{z}\right).
\end{align}
We can express the resulting right-handed electron asymmetry as
\begin{align}
    \label{Yekappadef}
    Y_{\Delta e}^{\rm{end}}=\tilde{\kappa}(z')Y_{\Delta e}^{\rm{eq}}(z'),
\end{align}
in full analogy with Eq.~\eqref{YDeltaXFinal}. Written in this form, with our abuse of notation for $\Lambda$, the results for $\tilde{\kappa}$ are equal to $\kappa$, which are displayed in Fig.~\ref{fig:parameter_space}. The parameter $\tilde{\kappa}$ 
indirectly parametrizes the outcome of WILG in the strong wash-in regime, which takes the following form in the scenario considered in this section
\begin{align}
    \label{eq:WILGOutcome}
    \frac{q_{B-L}^{\rm{win}}}{s}=x_{e}(z_{B-L})\tilde{\kappa}(z')Y_{\Delta e}^{\rm{eq}}(z').
\end{align}
In the scenario where the portal interaction attains thermal equilibrium, we have 
$Y_{\Delta e}^{\rm{eq}}(z')\sim Y_{Xe}\sim Y_{\Delta X}(z')$. This condition implies that, if the BAU is explained via WILG, the ADM mass must lie in the range $\mathcal{O}(1)-\mathcal{O}(10)$ GeV. In the FI regime, where $\tilde{\kappa}\ll 1$, the final ADM yield is typically much larger than the BAU, and one would expect ADM masses in the range $\mathcal{O}(\tilde{\kappa})-\mathcal{O}(10\,\tilde{\kappa})$ GeV. Specifically, the black dashed line in Fig.~\ref{fig:parameter_space} corresponds to DM masses in the MeV range. 

Finally, we note that since Asymgenesis builds on the type-I seesaw extension of the SM, it might be interesting to apply the framework to BSM models with gauged $U(1)_{B-L}$ symmetry. Contrary to conventional ADM models, Asymgenesis can accommodate scenarios where the charge-transfer mechanism decouples at temperatures that are orders of magnitude above the $B-L$-breaking scale, and still result in comparable abundances of baryonic matter and ADM.

%%%%%%%%%%%%%%%%%%%%%%%%%%%%%%%%%%%%%%%%%%%%%%%%%%%%%%%%%%%%%%%%%%%%%%%%%%%%%%%%%%%%%%%%%%%%%%%%%%%%
%%%%%%%%%%%%%%%%%%%%%%%%%%%%%%%%%%%%%%%%%%%%%%%%%%%%%%%%%%%%%%%%%%%%%%%%%%%%%%%%%%%%%%%%%%%%%%%%%%%%
%%%%%%%%%%%%%%%%%%%%%%%%%%%%%%%%%%%%%%%%%%%%%%%%%%%%%%%%%%%%%%%%%%%%%%%%%%%%%%%%%%%%%%%%%%%%%%%%%%%%

\noindent \textbf{Conclusion}\,---\,In this work, we have proposed \textit{Asymgenesis}, a framework based on the type-I seesaw model that can be used to account for the BAU and the observed DM density from initial conditions appropriate to either ADM or WILG. Compared to standard ADM scenarios that rely on higher-dimensional operators, Asymgenesis offers significantly greater flexibility in the charge-transfer interactions between the dark and visible sectors. Crucially, it disentangles the temperature at which the charge-transfer interactions decouple from the temperature scale at which $B-L$ violation becomes effective. By combining the advantages of WILG and ADM, the Asymgenesis framework considerably widens the model-building landscape for asymmetric dark matter.

%%%%%%%%%%%%%%%%%%%%%%%%%%%%%%%%%%%%%%%%%%%%%%%%%%%%%%%%%%%%%%%%%%%%%%%%%%%%%%%%%%%%%%%%%%%%%%%%%%%%

\medskip\noindent
\textit{Acknowledgments}\,---\,We are grateful to Kai Schmitz for insightful discussions and for pointing out an incorrect factor in Eq.~(3) in the first arXiv version. We thank Mathias Becker and Nicholas Leister for comments on the draft. M.\,A.\,M.\ acknowledges support from the Deutsche Forschungsgemeinschaft (DFG) Collaborative Research Centre ``Neutrinos and Dark Matter in Astro- and Particle Physics'' (SFB 1258).
The work of M.\,A.\,M.\ and S.\,W.\ was supported by the Cluster of Excellence “Precision Physics, Fundamental Interactions, and Structure of Matter” (PRISMA$^+$ EXC 2118/1) funded by the Deutsche Forschungsgemeinschaft (DFG, German Research Foundation) within the German Excellence Strategy (Project No. 390831469).

%%%%%%%%%%%%%%%%%%%%%%%%%%%%%%%%%%%%%%%%%%%%%%%%%%%%%%%%%%%%%%%%%%%%%%%%%%%%%%%%%%%%%%%%%%%%%%%%%%%%
\small

\bibliographystyle{JHEP} 
\bibliography{PLB2}
%%%%%%%%%%%%%%%%%%%%%%%%%%%%%%%%%%%%%%%%%%%%%%%%%%%%%%%%%%%%%%%%%%%%%%%%%%%%%%%%%%%%%%%%%%%%%%%%%%%%

\newpage
\onecolumngrid
\newpage

%%%%%%%%%%%%%%%%%%%%%%%%%%%%%%%%%%%%%%%%%%%%%%%%%%%%%%%%%%%%%%%%%%%%%%%%%%%%%%%%%%%%%%%%%%%%%%%%%%%%

\appendix

\section{On the relation between Asymgenesis and other charge-transfer mechanisms}
\label{Appendix}
In the main part of this letter, we have focused on the particularly simple scenario in which $\mathcal{O}_{C}^{\rm{(SM)}}$ in Eq.~\eqref{eq:operator} is a Lorentz scalar, gauge invariant, renormalizable, and violates the SM charge $C$. However, the Asymgenesis framework does not require any of these conditions to be satisfied. The purpose of this Appendix is to illustrate how well-known charge-transfer mechanisms are recovered once these conditions are relaxed, and explain how Asymgenesis provides a framework to generalize these mechanisms. In the rest of the discussion, we will allow $\mathcal{O}_{C}^{\rm{(SM)}}$ to transform nontrivially under Lorentz transformations and assume that $\mathcal{O}_X^{(D)}$ can be chosen such that $\mathcal{O}_{C}^{\rm{(SM)}}\mathcal{O}_X^{(D)}/\Tilde{\Lambda}^n$ is a Lorentz scalar. This is already sufficient to recover operators based on $\mathcal{O}_{C}^{\rm{(SM)}}=(LH)$~\cite{Ibe:2011hq}, which breaks lepton number. Below, we relax the remaining assumptions one by one. 

\textbf{Relaxing renormalizability:} To recover standard transfer operators that violate baryon number, such as the neutron portal operator $(u_Rd_Rd_R)$, one has to allow  $\mathcal{O}_{C}^{\rm{(SM)}}$ to be nonrenormalizable, since no renormalizable SM operator violates $B$. This also allows one to recover more lepton-number violating operators, such as $(LH)^2$~\cite{Cohen:2009fz,Ibe:2011hq}.

\textbf{Relaxing gauge invariance:} This corresponds to allowing the dark sector to contain a particle that carries SM gauge charge. For example, there are constructions considering operators of the form $X^2H^2$~\cite{Servant:2013uwa}, where $X$ is a fermionic doublet under $SU(2)_L$, and $\chi^2H^n$~\cite{Boucenna:2015haa}, where $\chi$ is an $SU(2)_L$ multiplet. These operators are, however, not recoverable unless the charge-violation condition on \(\mathcal{O}_{C}^{\rm (SM)}\) is also relaxed; see the discussion below. Nevertheless, they serve to illustrate how Asymgenesis can be used to generalize such constructions. In particular, one may simply replace the SM Higgs field with a fermion bilinear composed of
a lepton (or quark) doublet and a corresponding right-handed singlet, for example \(H \rightarrow \bar e_R L\) (or \(H \rightarrow \bar d_R Q\)).

\textbf{Relaxing charge-violation condition on $\mathcal{O}_{C}^{\rm{(SM)}}$:} So far, we have not recovered existing transfer mechanisms that do not violate $B$, $L$, or $B-L$, notably the Higgsogenesis mechanism~\cite{Servant:2013uwa} (see also~\cite{Blum:2012nf}). Higgsogenesis can be recovered by relaxing the assumption that $\mathcal{O}_{C}^{\rm{(SM)}}$ violates a SM charge, and as we will argue below, the Asymgenesis framework allows a substantial generalization of the Higgsogenesis mechanism. 

For completeness, we briefly review the main idea underlying Higgsogenesis: A conserved global charge in the thermal plasma enforces nonzero chemical potentials for other particle species through fast interactions. If there is a nonzero $B-L$ asymmetry in the thermal plasma, Yukawa interactions can bias a Higgs asymmetry between $H$ and $H^\dagger$. The Higgs asymmetry can then be transferred to the dark sector through a transfer operator. The inverse process where a primordial asymmetry produced in the dark sector could be transferred to the visible sector, is also possible. However, a successful application of the inverse process requires further assumptions, such as e.g., the electroweak phase transition (EWPT) to be first order to avoid devastating wash out of the baryon asymmetry, see~\cite{Servant:2013uwa} for details. 

Asymgenesis can also be applied with the transfer operator used in Higgsogenesis. For asymmetry transfer from the visible to the dark sector, Higgsogenesis assumes a primordial nonzero $B-L$ asymmetry. Asymgenesis, by contrast, does not require a nonzero $B-L$ asymmetry at any point while charge is transferred from the visible to the dark sector. In fact, Asymgenesis allows a generalization of Higgsogenesis where the primordial $B-L$ charge used in Higgsogenesis can be replaced with any of the (approximately) conserved SM charges, as the latter can also induce a nonzero chemical potential for the Higgs. Meanwhile, for charge transfer from the dark sector to the SM, Ref.~\cite{Servant:2013uwa} does not require $B-L$ violation, but instead relies on the EWPT to be strongly first order. Instead of requiring the EWPT to be strongly first-order for successful baryogenesis from a primordial dark asymmetry, Asymgenesis requires the existence of right-handed neutrinos.

A schematic of differences between standard ADM, Higgsogenesis, and Asymgenesis is summarized in Fig.~\ref{fig:HiggsADMAsym}. The figure emphasizes the defining feature of Asymgenesis: It can operate successfully without $B-L$-violation in the charge transfer between the visible and dark sectors, and does not require the SM plasma to be $B-L$-asymmetric at the time when charge-transfer processes are efficient. 

\begin{figure}
\begin{center}
\begin{tikzpicture}[>=stealth, scale=1.2]

\node (A) [
    draw,
    minimum width=6cm,
    minimum height=1cm,
    text width=5.6cm,        % controls wrapping
    align=center,          % centers multi-line text
    inner sep=2mm          % padding between text and border
  ] {Asymgenesis};

\node (B) [
    draw,
    minimum width=6cm,
    minimum height=1cm,
    text width=5.6cm,
    align=center,
    inner sep=2mm,
    right=-0.1mm of A   
  ] {Standard ADM};

\node (C) [
    draw,
    minimum width=6cm,
    minimum height=1cm,
    text width=5.6cm,
    align=center,
    inner sep=0mm,
    below=-0.1mm of A   
  ] {Asymgenesis (for example generalized Higgsogenesis)};

\node (D) [
    draw,
    minimum width=6cm,
    minimum height=1cm,
    text width=5.6cm,
    align=center,
    inner sep=2mm,
    below=-0.1mm of B   
  ] {Higgsogenesis};

\node (E) [
    minimum width=6cm,
    minimum height=1cm,
    text width=5.6cm,
    align=center,
    inner sep=2mm,
    left=-0.1mm of A  
  ] {Charge transfer violates a \\
  conserved SM charge $C$};

\node (F) [
    minimum width=6cm,
    minimum height=1cm,
    text width=5.6cm,
    align=center,
    inner sep=2mm,
    left=-0.1mm of C  
  ] {Charge transfer does not\\ violate any conserved SM charges $C$};

\node (G) [
    minimum width=6cm,
    minimum height=1cm,
    text width=5.6cm,
    align=center,
    inner sep=2mm,
    above=+0.1mm of A  
  ] {May use a charge $C\neq B\!-\!L$};

\node (G) [
    minimum width=6cm,
    minimum height=1cm,
    text width=5.6cm,
    align=center,
    inner sep=2mm,
    above=+0.1mm of B  
  ] {$C=B\!-\!L$};

\end{tikzpicture}
\end{center}
\caption{Schematic comparison between Asymgenesis, standard ADM based on charge transfer via higher-dimensional operators, and Higgsogenesis. See text for details.}
\label{fig:HiggsADMAsym}
\end{figure}

\end{document}